\documentclass[12pt]{article}

\usepackage{amsmath}
\usepackage{amssymb}
\usepackage{tikz-cd}
\usepackage{setspace,enumitem,soul,color,hyperref}

\usepackage{authblk}
\author[1,2,3]{Li-Chun Zhang}
\affil[1]{\em \small University of Southampton (email: L.Zhang@soton.ac.uk)}
\affil[2]{\em \small Statistics Norway}
\affil[3]{\em \small University of Oslo}
\title{\LARGE Sampling designs for epidemic prevalence estimation}
\date{}

\begin{document}

\maketitle

\begin{abstract}
Intuitively, sampling is likely to be more efficient for prevalence estimation, if the cases (or positives) have a relatively higher representation in the sample than in the population. In case the virus is transmitted via personal contacts, contact tracing of the observed cases (but not noncases), to be referred to as \emph{adaptive network tracing}, can generate a higher yield of cases than random sampling from the population. The efficacy of relevant designs for cross-sectional and change estimation is investigated.  The availability of these designs allows one unite tracing for combating the epidemic and sampling for estimating the prevalence in a single endeavour.  
\end{abstract}

\noindent \textit{Key words:} case network, adaptive cluster sampling, panel, graph sampling

\section{Introduction}

Let the population $U$ be of size $N$. Let $y_i =1$ if person $i\in U$ is a \emph{case}, and $y_i =0$ otherwise. Let the population case \emph{total} and \emph{prevalence} be given by, respectively,  
\[
\theta = \sum_{i\in U} y_i  \qquad\text{and}\qquad \mu = \theta/N
\]
For the most efficient design of a sample $s$ that is of the size $\theta$, one should include all the cases with probability $\pi_i =1$ and all the noncases with probability $\pi_i =0$, so that the Horvitz-Thompson estimator would have zero error and zero sampling variance, i.e.
\[
\hat{\theta}_{HT} = \sum_{i\in s} y_i/\pi_i = \sum_{\substack{i\in U\\ y_i=1}} y_i \equiv \theta \quad\text{and}\quad V(\hat{\theta}_{HT}) = 0
\]
Although the ideal is unachievable in reality, it does echo the intuition that one should aim at higher representation of cases in the sample than in the population. 

To increase the sample proportion of cases in practice, one may select an initial sample, denoted by $s_0$, and include additional units by repeatedly tracing all the contacts of the cases, starting from $s_0$ and until no more cases can be obtained by tracing. Such contact tracing is said to be \emph{adaptive} since it is implemented only to cases but not noncases. Moreover, one may aim at $\mbox{Pr}(i\in s_0 | y_i =1) > \mbox{Pr}(i\in s_0 | y_i =0)$ when selecting the initial sample $s_0$, i.e. \emph{size-biased} initial sampling provided relevant information is available.   

Below we consider designs using adaptive network tracing and size-based initial sampling for both cross-sectional and change estimation of prevalence.

\section{Cross-sectional designs}

Under network sampling (Sirken, 1970), ``siblings report each other'' are needed to reach a ``network'' of siblings following an initial sample of households. Under adaptive cluster sampling (ACS, Thompson, 1990), the final sample depends on the ``network'' relationship among the units as well as the values of the surveyed units. Combining the two, we define a \emph{case network} to be a set of cases that can be connected to each other via personal contacts, and adaptive network tracing as repeated contact tracing from any case until its case network is exhausted. In the terminology of ACS, any noncase that is in contact with a case will be referred to as a (network) \emph{edge node}. Due to the adaptive nature of tracing, one would observe an edge node, if its in-contact case network intersects $s_0$; but an edge node in $s_0$ would not lead to the observation of its in-contact case network.

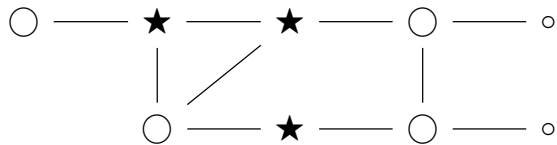
\begin{figure}[ht]
\centering
\begin{tikzcd}
\bigcirc &  \bigstar \arrow[l, dash] \arrow[r, dash] & \bigstar \arrow[r, dash] & \bigcirc & \circ \arrow[l, dash] \\
& \bigcirc \arrow[u, dash] \arrow[r, dash] \arrow[ur, dash] & \bigstar \arrow[r, dash] & \bigcirc \arrow[u, dash] & \circ \arrow[l, dash]
\end{tikzcd} \\
\caption{Cases $\bigstar$, noncase edge nodes $\bigcirc$, other noncases $\circ$} \label{fig:network}
\end{figure}

One can represent the individuals and their contacts as a \emph{population graph}, denoted by $G = (U, A)$, where the edge set $A$ contains all the relevant contacts. We shall consider $G$ to be undirected and simple, i.e. $(ij), (ji)\in A$ iff $i$ and $j$ are in-contact. Adaptive network tracing from an initial probability sample here is a special form of ACS, where $y_i$ is binary and the noncases with $y_i$ contributes nothing to the population total $\theta$; and ACS is a special case of graph sampling (Zhang and Patone, 2017). An illustration of two case networks, their edge nodes and other noncases in $G$ is given in Figure \ref{fig:network}.

Given an initial sample $s_0$ of size $m$ from $U$. For each case $i$ in $s_0$ with $y_i =1$, include all the adjacent individuals (in $G$), denoted by $\nu_i = \{ j : (ij) \in A\}$. Let 
\[
s_1 = \bigcup_{i\in s_0} \nu_i \setminus s_0
\]
be the 1st-wave sample of individuals. Repeat the above observation procedure for everyone in $s_1$, which may or may not result in a non-empty 2nd-wave sample 
\[
s_2 = \bigcup_{i\in s_1} \nu_i \setminus (s_0\cup s_1)
\]
The observation procedure is repeated, till it results in an empty wave sample. The final sample is given by $s = \bigcup_{r=0}^q s_r$, where the $q$th-wave sample is empty.

The sample $s$ resulting from the ACS can be divided into three parts: (i) a set of case networks, where those in the same network all have $y_i = 1$ and are connected to each other via the edges in $A$, (ii) the \emph{edge units}, where an edge unit has $y_i =0$ and is adjacent (in $G$) to at least one sampled network of cases, and (iii) the remaining individuals in the initial sample $s_0$ who have $y_i =0$ and do not belong to (i) and (ii).

The sample inclusion probability of a case $i$ in network $\kappa$ is the probability that the intersection of $s_0$ and $\kappa$ is non-empty. This follows from the fact that all the cases in the same network are observed together iff any of them is observed. The sample inclusion probability of an edge node cannot be calculated, as long as it is unknown whether there are other out-of-sample adjacent networks. \textit{However, this does not matter here, since the corresponding unit has $y_i \equiv 0$ and its contribution to the estimator of $t$ is 0 regardless its inclusion probability.} The same goes for the sampled individuals in part (iii). 

The Horvitz-Thompson (HT) estimator of $Y$ is given by
\begin{equation} \label{ACS-HT}
\hat{\theta}_{HT}^{ACS} = \sum_{i\in s} \frac{y_i}{\pi_{(i)}} = \sum_{\kappa} \frac{n_{\kappa}}{\pi_{(\kappa)}}
\end{equation}
where $\kappa$ is a case network of size $n_{\kappa}$ and $\pi_{(\kappa)}$ is its inclusion probability under ACS, which is the same as $\pi_{(i)}$ for any $i\in \kappa$. Notice the distinction between $\pi_{(i)} = \mbox{Pr}(i\in s)$ and the initial sample inclusion probability $\pi_i = \mbox{Pr}(i\in s_0)$. Since a case may be included in $s$ via its network, even when it is not directly selected in $s_0$, we have
\[
\pi_{(i)} > \pi_i
\]
as long as the network of case $i$ contains one or more other cases. This means adaptive network tracing can increase the yield of cases in $s$ compared to only using $s_0$. The sampling variance of $\hat{\theta}_{HT}^{ACS} $ is given in the Appendix.

\subsection{Some simulation results}

Let $\eta$ be the odds of case selection in the initial sample $s_0$, which is defined as the ratio of the probability that a case is included in $s_0$ against that of a non-case. We have positively size-biased initial sampling if $\eta >1$. 

\begin{table}[ht]
\begin{center}
\caption{RE of ACS given equal-size ($k$) case networks in population of size $N = 10^5$, prevalence $\mu =0.01$ and total $\theta = 10^3$. Initial sample of size $m$ by SRS ($\eta=1$) or size-biased sampling ($\eta =2$), ACS with sample size $n = |s|$ by adaptive network tracing.}
\begin{tabular}{rc | rcc | rcc | rcc} \hline \hline
\multicolumn{2}{c}{SRS ($\eta=1$)} & \multicolumn{3}{|c|}{ACS, $k=100$} & \multicolumn{3}{c}{ACS, $k=10$} 
& \multicolumn{3}{|c}{ACS, $k=2$} \rule[-1.0ex]{0pt}{3.5ex} \\
$m$ & CV & $E(n)$ & CV & RE & $E(n)$ & CV & RE & $E(n)$ & CV & RE \\ \hline
1000 & 0.31 & 1631 & 0.24 & 0.58 & 1085 & 0.31 & 0.96 & 1010 & 0.31 & 0.99 \\
1630 & 0.24 & 2423 & 0.15 & 0.40 & 1766 & 0.24 & 0.93 & 1646 & 0.24 & 0.99 \\
2420 & 0.20 & 3306 & 0.10 & 0.23 & 2614 & 0.19 & 0.89 & 2443 & 0.20 & 0.99 \\
5000 & 0.14 & 5944 & 0.02 & 0.03 & 5352 & 0.12 & 0.79 & 5048 & 0.14 & 0.97 \\
10000 & 0.09 & 10900 & 0.00 & 0.00 & 10551 & 0.07 & 0.59 & 10090 & 0.09 & 0.95 \\ \hline \hline
\multicolumn{2}{c}{Size-biased ($\eta=2$)} & \multicolumn{3}{|c|}{ACS, $k=100$} & \multicolumn{3}{c}{ACS, $k=10$} 
& \multicolumn{3}{|c}{ACS, $k=2$} \rule[-1.0ex]{0pt}{3.5ex} \\
$m$ & CV & $E(n)$ & CV & RE & $E(n)$ & CV & RE & $E(n)$ & CV & RE \\ \hline
1000 & 0.22 & 1840 & 0.13 & 0.32 & 1160 & 0.21 & 0.91 & 1020 & 0.22 & 0.99 \\
5000 & 0.09 & 5901 & 0.00 & 0.00 & 5549 & 0.07 & 0.60 & 5090 & 0.09 & 0.95\\
10000 & 0.06 & 10802 & 0.00 & 0.00 & 10692 & 0.04 & 0.31 & 10159 & 0.06 & 0.89 \\ \hline \hline
\end{tabular} \label{tab:pers}
\end{center} \end{table}

Table \ref{tab:pers} presents some results for the relative efficiency (RE) of ACS, defined as the ratio of the variance of the HT estimator under ACS against that based on the initial sample $s_0$, which is either selected by SRS ($\eta = 1$) or sized-biased sampling with $\eta = 2$. All the cases in the population are divided into networks, which have the same size $k$. 

It can be seen that ACS is increasingly more efficient than SRS as $m$ increases, if one compares the CVs of SRS with $m$ and ACS with $E(n)$, where $m \approx E(n)$. The gain is more pronounced given large networks, e.g. $k = 100$.
Given initial SRS of size $10^4$, ACS requires about 900 extra individuals, by which the sampling variance is reduced to 0.00, because any case network intersects $s_0$ almost certainly. The reduction is quicker given initial size-biased sampling, where e.g. the variance is already 0.00 at $m = 0.5\times 10^4$.

Unsurprisingly, ACS has basically no gains given only small case networks with $k = 2$, where size-biased sampling would be the chief means for reducing variance, e.g. with $\eta=2$ the variance of the initial sample estimator is about halved given any $m$ in Table \ref{tab:pers}.   

Together, size-biased sampling and adaptive network tracing can enhance each other, generating extra gains when they are applied jointly. 

Finally, note that the inclusion probabilities $\pi_{(\kappa)}$ and $\pi_{(\kappa\ell)} = \mbox{Pr}(\kappa\in s, \ell\in s)$ are easy to compute under initial SRS. But these sample inclusions probabilities are usually unknown given unequal probability initial sampling,. When the sampling fraction is not too high, it is convenient to treat the initial sampling as if it were Poisson sampling, where the individuals are independently selected. It has been verified empirically that the approximation holds well in the simulation settings here, including the highest sampling fraction 10\%.

\subsection{Other designs}

The observation procedure of ACS is \emph{network exhaustive}, in the sense that a network of cases are observed altogether if any of them is selected in $s_0$. There could be an issue if a network is too large to be surveyed completely due to practical reasons. If it is possible to measure $\psi_{ij}$ as the \emph{strength} of $(ij)\in A$, then one may define adaptively the eligible adjacency of $i$ to be $\nu_i^* = \{ j : (ij)\in A, \psi_{ij} > \psi_0 \}$ for a chosen threshold $\psi_0$, and include $\nu_i^*$ if $y_i = 1$. The resulting method may be referred to as \emph{doubly adaptive cluster sampling (DACS)}, since it is based on two threshold values $y_i >0$ and $\psi_{ij} > \psi_0$.   

Imposing a maximum number of waves, say $q$, is another way to curtail large networks. The sampling will be terminated after the $q$-th wave, even if $s_q \neq \emptyset$, yielding the sample $s = \sum_{r=0}^q s_r$. This is a \emph{$q$-stage adaptive snowball sampling (qASBS)} design, given the threshold $y_i >0$, or $\psi_{ij} > \psi_0$ in addition. Unlike ACS or DACS, \emph{not} all the observed network cases can be used for estimation, but only those whose ancestors under qASBS have already been observed. This is because the observation procedure is not network exhaustive here, even though it is if $q = \infty$. 

Both DACS and qASBS can be handled as bipartite incidence graph sampling, using the methods of Zhang and Oguz-Alper (2020). We do not pursue the details here.  

Finally, stratified multistage sampling (e.g. Cochran, 1972) is a reliable standard design approach, though it is likely to be less efficient than ACS and its modified versions. One may consider a combined approach at the onslaught of an epidemic, where an ongoing national survey by stratified multistage sampling can be redeployed . 
\begin{itemize}[leftmargin=6mm]
\item Change the content of the ongoing sample survey to secure a baseline precision for the estimation of prevalence. Denote by $s'$ this sample by stratified multistage design.
\item Apply ACS, or its modified version, to a subsample $s_0$ of $s'$, to collect additional data about cases and contact relationships. E.g. $s_0$ may consist of the asymptomatic cases.   
\item For the estimation of $\theta$ and $\mu$, one may combine the two samples. The data collected under ACS are useful for epidemiological modelling and analysis and, if needed, for providing the necessary evidence for a greater rollout of ACS afterwards. 
\end{itemize}

\subsection{Population of households} 

Households can be envisaged as social bubbles, within which contact is hard to avoid. Testing is likely to be more cost-effective when administered to the whole household than just a single member of it. It is possible that the whole household will be placed under quarantine, once any of them is tested positive. For all these reasons one may treat each household as a single unit. Denote by $\mathbb{G} = (\mathbb{H}, \mathbb{A})$ the population graph, where $\mathbb{H}$ consists of all the households, and $\mathbb{A}$ the contacts between any two households via their members. Size-biased sampling and adaptive network tracing in $\mathbb{G}$ follow the same definition as in $G = (U, A)$, but the actual design effects will differ between the two set-ups.

\begin{table}[ht]
\begin{center}
\caption{RE of ACS given equal-size ($k$) case networks in household population of size $N = 10^5$, prevalence $\mu =0.01$ and total $\theta = 10^3$. Initial sample of size $m$ by SRS ($\eta=1$) or size-biased sampling ($\eta =2$), ACS with sample size $n = |s|$ by adaptive network tracing.}
\begin{tabular}{rc | rcc | rcc | rcc} \hline \hline
\multicolumn{2}{c}{SRS ($\eta=1$)} & \multicolumn{3}{|c|}{ACS, $k=100$} & \multicolumn{3}{c}{ACS, $k=10$} 
& \multicolumn{3}{|c}{ACS, $k=2$} \rule[-1.0ex]{0pt}{3.5ex} \\
$m$ & CV & $E(n)$ & CV & RE & $E(n)$ & CV & RE & $E(n)$ & CV & RE \\ \hline
1000 & 0.35 & 1628 & 0.24 & 0.45 & 1086 & 0.31 & 0.77 & 1010 & 0.34 & 0.89 \\
5000 & 0.16 & 5944 & 0.02 & 0.03 & 5351 & 0.13 & 0.63 & 5048 & 0.14 & 0.87 \\
10000 & 0.11 & 10900 & 0.00 & 0.00 & 10552 & 0.07 & 0.49 & 10090 & 0.10 & 0.84 \\ \hline \hline
\multicolumn{2}{c}{Size-biased ($\eta=2$)} & \multicolumn{3}{|c|}{ACS, $k=100$} & \multicolumn{3}{c}{ACS, $k=10$} 
& \multicolumn{3}{|c}{ACS, $k=2$} \rule[-1.0ex]{0pt}{3.5ex} \\
$m$ & CV & $E(n)$ & CV & RE & $E(n)$ & CV & RE & $E(n)$ & CV & RE \\ \hline
1000 & 0.25 & 1844 & 0.13 & 0.26 & 1162 & 0.22 & 0.71 & 1019 & 0.23 & 0.86 \\
5000 & 0.11 & 5901 & 0.00 & 0.00 & 5547 & 0.08 & 0.47 & 5089 & 0.10 & 0.85\\
10000 & 0.07 & 10802 & 0.00 & 0.00 & 10691 & 0.04 & 0.25 & 10159 & 0.06 & 0.80 \\ \hline \hline
\end{tabular} \label{tab-hsh}
\end{center} \end{table}
 
Table \ref{tab-hsh} presents some results for the RE of ACS based on adaptive network tracing, which are comparable to those of Table \ref{tab:pers}. The only difference is that sampling and network tracing are from a population of households instead of persons. The stipulated household size distribution is $(0.38, 0.30, 0.12, 0.20)$ for household size $(1, 2, 3, 4)$, among both the households of cases and the households of non-cases. The differences of $E(n)$ in Table \ref{tab:pers} and \ref{tab-hsh} reflect the magnitude of Monte Carlo simulation error in these results.

It can be seen that the variances are larger than the corresponding ones (Table \ref{tab:pers}) under sampling of individuals. However, the increases under ACS are smaller, such that the relative efficiency gains by adaptive network tracing are actually increased compared to sampling of persons, where the RE is appreciable even when the networks are small, e.g. $k = 2$. The relative efficiency of size-biased initial sampling is similar to that in Table \ref{tab:pers}.

\section{Designs for estimating changes}

The population graph is dynamic over time, denoted by $G_t = (U_t, A_t)$, for time point $t=1, 2, ...$ Even when $U_t$ is fixed, the contact edge set $A_t$ that are relevant for the time point $t$ will be dynamic, unless the society is a state of strict lockdown. In addition to fixing $U_t = U$, they are at least two ways of allowing it to be dynamic. 
\begin{itemize}[leftmargin=6mm]
\item[a.] Let $U_t$ be the union of a fixed $U$ and those that can be linked to it via $A_t$.   
\item[b.] One can define the population recursively, where $U_t$ is the union of $U_{t-1}$ and those that can be linked to it via $A_t$, starting from $U_1 = U$.
\end{itemize}
Insofar as it is possible to estimate $N_t = |U_t|$ when the population is treated as dynamic, one may allow the definition of $U_t$ to depend on the circumstances in applications.
We consider three designs for change estimation.
\begin{itemize}[leftmargin=4mm]
\item In the \emph{panel} design, the sample once selected is fixed over time. This accommodates only the fixed target population $U_t = U$.  

When the change  $\mu_{t'} - \mu_t$ is estimated based on two independent samples at $t$ and $t'$, respectively, the variance is the sum of those of $\hat{\mu}_t$ and $\hat{\mu}_{t'}$. The variance is reduced, if the panel design induces a positive correlation between $\hat{\mu}_t$ and $\hat{\mu}_{t'}$. Suppose that based on two independent samples of the same size $n$, the sampling variance of the change estimator (between two proportions) is
\[
V_{\mbox{ind}}(\hat{\mu}_{t'} - \hat{\mu}_t) = V(\hat{\mu}_{t'}) + V(\hat{\mu}_t) 
= \frac{1}{n} \big( \mu_{t'} (1- \mu_{t'}) + \mu_t (1- \mu_t) \big) \doteq \frac{1}{n} \big( \mu_{t'} + \mu_t \big)
\]
as long as $1- \mu_{t'} \doteq 1$ and $1- \mu_t \doteq 1$. Suppose $N_{t'} =N_t$ and $\mu_{t'} = \mu_t$ for simplicity. Let $\lambda_+$ be the proportion of the new cases between $t$ and $t'$, and $\lambda_-$ that of the closed cases, where  $\lambda_+ \leq \mu_{t'}$ and $\lambda_- \leq \mu_t$ by definition. Let $d_i = 0$ if individual $i$ has no change of case status, $d_i=1$ if $i$ becomes a case, and $d_i =-1$ if $i$ becomes a closed case, such that $\mu_{t'} - \mu_t$ is the population mean of $d_i$. Based on a panel of size $n$, it can be shown that
\[
V_{\mbox{pnl}}(\hat{\mu}_{t'} - \hat{\mu}_t) \doteq \frac{1}{n} (\lambda_+ + \lambda_-) 
\leq \frac{1}{n} \big( \mu_{t'} + \mu_t \big) = V_{\mbox{ind}}(\hat{\mu}_{t'} - \hat{\mu}_t)  
\]

\item Under the \emph{panel ACS (pACS)} design, only the initial sample $s_0$ is fixed over time, but the sample $s(t)$ obtained based on $s_0$ and $A_t$ via adaptive network tracing will vary, as both the case status $y_{i,t}$ for individual $i$ at time $t$ and the population contacts $A_t$ change over time. In addition to fixed population $U_t = U$, the pACS design allows for dynamic population definition (a) above. The estimator of change can be given as the difference between the corresponding HT-estimators \eqref{ACS-HT} at two time points. 

\item Under the \emph{iterated ACS (iACS)} design, let $s(t)$ be the sample by ACS based on $s_0$ and $A_t$ for time $t$, and let $s(t+1)$ again be given by ACS but based on $s_0 = s(t)$ and $A_{t+1}$ for time $t+1$. In addition to fixed population $U_t = U$, the iACS design allows for both the dynamic population definitions (a) and (b) above.

Since $s_0 \subseteq s(t)$, the sample $s(t+1)$ can only be larger under iACS than under pACS, while the sample $s(t)$ is the same under both. As explained later, the HT-estimator estimator for $\theta_{t+1}$ is unavailable under iACS, so that a different unbiased estimator is used. Simulations later show that neither design dominates the other generally.  
\end{itemize}

Finally, as the testing capacity increases, one may wish to \emph{boost the sample} over time. Let $s_b(t)$ be the supplement sample selected at time $t$, where $s_b(t) = \emptyset$ if no supplement sample is added. It can be obtained by ACS, via an initial sample $s_{0,t}$ selected at time $t$. One only needs to keep track of the inclusion probabilities $\pi_i$ and $\pi_{ij}$ for $i\neq j\in s_0 \bigcup_{r=1}^t s_{0,r}$, which is possible if all the initial samples $s_{0,t}$ are drawn from a fixed frame $U$.

\subsection{Change estimators} \label{change}

The HT estimator of change $\nabla_{t,t+1} = \mu_{t+1} - \mu_t$ under the panel design is given by 
\[
\hat{\nabla}_{t,t+1}^{panel} = \frac{1}{N} \sum_{i\in s_0} \frac{1}{\pi_i} (y_{i,t+1} - y_{i,t})
\]
where $\pi_i$ is the inclusion probability of $i\in s_0$, and $s_0$ is the panel. 

The HT estimator of $\nabla_{t,t+1}$ under the panel ACS design is given by 
\[
\hat{\nabla}_{t,t+1}^{pACS} = \frac{1}{N_{t+1}} \sum_{i\in s(t+1)} \frac{y_{i,t+1}}{\pi_i(t+1)} 
- \frac{1}{N_t} \sum_{i\in s(t)} \frac{y_{i,t}}{\pi_i(t)}
\]
where $s(t)$ is the sample at time $t$ by ACS based on $s_0$ and $A_t$, and $\pi_i(t)$ is the inclusion probability of $i\in s(t)$, and similarly for $s(t+1)$ and $\pi_i(t+1)$. That it, one applies \eqref{ACS-HT} at each time point and take the difference between them. The sampling variance of $\hat{\nabla}_{t,t+1}^{pACS}$ is given in the Appendix. It is possible to plug in the HT-estimates of $\hat{N}_t$ and $\hat{N}_{t+1}$ in the above, if one adopts the dynamic population definition (a). 

Under the iterated ACS design, an unbiased estimator of $\nabla_{t,t+1}$ is given by
\begin{equation} \label{iACS}
\hat{\nabla}_{t,t+1}^{iACS} = \frac{1}{N_{t+1}} \Big( \sum_{\substack{i\in s(t)\\ y_{i,t}=1}} \frac{y_{i,t+1}}{\pi_i(t)} 
+ \sum_{\substack{i\in s(t)\\ y_{i,t}=0}} \frac{y_{i,t+1}}{\pi_i} \Big) - \frac{1}{N_t} \sum_{i\in s(t)} \frac{y_{i,t}}{\pi_i(t)}
\end{equation}
The two terms in the parentheses form an unbiased Hansen-Hurwitz (HH) type estimator of $\theta_{t+1}$, where the values $\{ y_{j,t+1}:j\in s(t+1)\}$ by iterated ACS based on $s(t)$ and $A_{t+1}$ are transformed to the constructed values $\{ z_i : i\in s(t) \}$. The estimator dates back to Birnbaum and Sirken (1965). Now that $y = 0$ or 1, we have $z_i =1$ using the multiplicity weights if $y_{i,t+1} =1$, and 0 otherwise, i.e. $z_i = y_{i,t+1}$. Meanwhile, the inclusion probability in $s(t)$ differs depending on whether an individual is case or not at $t$, corresponding to the two terms given above, respectively, since $s(t)$ is obtained by ACS based on $s_0$ and $A_t$. The sampling variance of $\hat{\nabla}_{t,t+1}^{iACS}$ is given in the Appendix. 
 
Notice that, under ACS, one can only calculate $\pi_i(t)$ and $\pi_{ij}(t)$ for $i, j\in s(t) \subset U_t$, but not any units out of $s(t)$. Since some units in $U_t\setminus s(t)$ may be included in the sample $s(t+1)$ under the iACS design, one generally cannot calculate the sample inclusion probabilities for all the units in the resulting $s(t+1)$. This is the reason why the HT-estimator for $\theta_{t+1}$ is generally unavailable under iACS. We refer to Patone and Zhang (2020) for a more extensive investigation as well as a synthesis these estimators.

The estimator $\hat{D}_{t,t+1}^{pACS}$ by panel ACS can be more efficient than $\hat{D}_{t,t+1}^{panel}$ under the panel design, because $s_0$ is a subsample of either $s(t)$ or $s(t+1)$, and ACS increases the sample inclusion probability of a case. Likewise between $\hat{D}_{t,t+1}^{iACS}$ and $\hat{D}_{t,t+1}^{panel}$. The RE between panel and iterated ACS designs is undetermined in general. On the one hand, the sample $s(t+1)$ based on $s_0$ and $A_{t+1}$ is a subsample of that based on $s_0 =s(t)$ and $A_{t+1}$ because $s_0 \subseteq s(t)$; on the other hand, the HH-type estimator of $\theta_{t+1}$ under iACS may be less efficient than the HT-estimator of $\theta_{t+1}$ under pACS. The strengths of the contrasting forces depend on how the case networks in $A_t$ and $A_{t+1}$ relate to each other.

\subsection{Simulation results over two time points}

New case networks may emerge from one time point to the next, whilst the existing ones may increase or decrease in their sizes. The speed may be quick or slow, at which a new case network emerges or an existing one grows or shrinks. Some settings over two time points are given in Table \ref{tab-2pop}, where both the population size and prevalence are constant, such that the target parameter is $\nabla_{1,2} =0$ in all the settings. Notice that the networks are all of size 2 at $t=1$ in the last three settings S1-S3. For the networks that are not growing, their sizes at $t=2$ are randomly assigned, subjected to the case total $\theta_2 =10^3$, such that some of them may simply disappear by chance.

\begin{table}[ht]
\begin{center}
\caption{Populations of constant size $N=10^5$ and case total $\theta = 10^3$ at $t=1,2$. With (number, size) of case networks: at $t=1$, $(\bar{\theta}, k)$ networks; at $t=2$, $(\bar{\theta}_+, k_+)$ or $(\bar{\theta}_-, k_-)$ existing networks of increasing or decreasing sizes, and $(\bar{\theta}', k')$ emerging networks.}
\begin{tabular}{c l cccc} \hline \hline
& & $t=1$ & \multicolumn{3}{c}{$t=2$} \\ \cline{4-6}
Setting & Characterisation & $(\bar{\theta}, k)$ & $(\bar{\theta}_+, k_+)$ & $(\bar{\theta}_-, k_-)$ 
& $(\bar{\theta}', k')$ \rule[-1.0ex]{0pt}{3.5ex} \\ \hline
L1 & Large, Quickly Evolving & (10, 100) & (2, 180) & (8, 80) & (0, 0) \\
L2 & Large, Quickly Emerging & (10, 100) & (0, 0) & (10, 80) & (2, 100) \\
L3 & Large, Slowly Emerging & (10, 100) & (0, 0) & (10, 90) & (5, 20) \\ \hline
M1 & Medium, Quickly Evolving & (100, 10) & (10, 46) & (90, 6) & (0, 0) \\
M2 & Medium, Quickly Emerging & (100, 10) & (0, 0) & (100, 6) & (10, 40) \\
M3 & Medium, Slowly Emerging & (100, 10) & (0, 0) & (100, 9) & (10, 10) \\ \hline
S1 & Small, Quickly Evolving & (500, 2) & (10, 42) & ($\leq$490, $\leq$2) & (0, 0) \\
S2 & Small, Quickly Emerging & (500, 2) & (0, 0) & ($\leq$500, $\leq$2) & (10, 40) \\
S3 & Small, Slowly Emerging & (500, 2) & (0, 0) & ($\leq$500, $\leq$2) & (50, 2) \\ \hline \hline
\end{tabular} \label{tab-2pop}
\end{center} \end{table}  
  
\begin{table}[h]
\begin{center}
\caption{Panel, pACS and iACS designs for settings in Table \ref{tab-2pop}.}
\renewcommand{\tabcolsep}{3mm}
\begin{tabular}{c ccc ccc ccc} \hline \hline
& \multicolumn{9}{c}{Initial SRS of Size $m =10^3$} \\ \cline{2-10}
(SE in $10^{-2}$) & L1 & L2 & L3 & M1 & M2 & M3 & S1 & S2 & S3 \\ \hline
SE($\hat{\nabla}_{t,t+1}^{panel}$) & 0.20 & 0.20 & 0.14 & 0.28 & 0.28 & 0.14 & 0.28 & 0.28 & 0.13 \\
RE($\hat{\nabla}_{t,t+1}^{pACS}$) & 0.71 & 0.73 & 0.90 & 0.89 & 0.89 & 0.98 & 0.89 & 0.91 & 0.98 \\
RE($\hat{\nabla}_{t,t+1}^{iACS}$) & 0.57 & 0.60 & 0.52 & 0.69 & 0.70 & 0.52 & 0.84 & 0.85 & 0.75 \\ \hline \hline
& \multicolumn{9}{c}{Initial SRS of Size $m =5\times 10^3$} \\ \cline{2-10}
(SE in $10^{-2}$) & L1 & L2 & L3 & M1 & M2 & M3 & S1 & S2 & S3 \\ \hline
SE($\hat{\nabla}_{t,t+1}^{panel}$) & 0.09 & 0.09 & 0.06 & 0.12 & 0.12 & 0.06 & 0.12 & 0.12 & 0.06 \\
RE($\hat{\nabla}_{t,t+1}^{pACS}$) & 0.09 & 0.11 & 0.38 & 0.62 & 0.63 & 0.87 & 0.65 & 0.64 & 0.91 \\
RE($\hat{\nabla}_{t,t+1}^{iACS}$) & 0.49 & 0.51 & 0.49 & 0.67 & 0.67 & 0.53 & 0.85 & 0.84 & 0.76 \\ \hline \hline
& \multicolumn{9}{c}{Initial Size-biased Sampling ($\eta=2$) of Size $m =10^3$} \\ \cline{2-10}
(SE in $10^{-2}$) & L1 & L2 & L3 & M1 & M2 & M3 & S1 & S2 & S3 \\ \hline
SE($\hat{\nabla}_{t,t+1}^{panel}$) & 0.17 & 0.17 & 0.12 & 0.24 & 0.24 & 0.12 & 0.24 & 0.24 & 0.12 \\
RE($\hat{\nabla}_{t,t+1}^{pACS}$) & 0.31 & 0.33 & 0.41 & 0.51 & 0.51 & 0.50 & 0.52 & 0.53 & 0.51 \\
RE($\hat{\nabla}_{t,t+1}^{iACS}$) & 0.70 & 0.69 & 0.68 & 0.79 & 0.81 & 0.68 & 0.89 & 0.90 & 0.84 \\ \hline \hline
\end{tabular} \label{tab-change}
\end{center} \end{table}  

\noindent
Table \ref{tab-change} presents some simulation results for the settings in Table \ref{tab-2pop}. The RE of an adaptive design is calculated against the panel design without adaptive network tracing.  
\begin{itemize}[leftmargin=4mm]
\item Overall, from top-right towards bottom-left in Table \ref{tab-change}, the RE of panel ACS is seen to improve quickly with the three initial values of sample size $m$, odds of case selection $\eta$ and case network size $k$. The RE of iterated ACS improves with $m$ except for S1-S3, but not with $\eta$, although it remains more efficient that the panel design as $\eta$ increases. The improvements are larger for panel ACS than iterated ACS.

\item For any given initial network size $k$, moving between the three patterns of case networks over time, the RE of panel ACS improves less for slowly than quickly changing networks, as $m$ increases. As the initial odds of case selection $\eta$ increases, the RE of the panel ACS become more uniform across all the patterns, .  

\item Given any combination of initial values of $(k, m, \eta)$, the RE of iterated ACS becomes more uniform across the three patterns of case networks, as $m$ and $\eta$ increase.

\item Between the two ACS designs, the panel ACS is more efficient given sufficiently large initial sample size $m$ and high odds of case selection $\eta$, whereas the iterated ACS is more efficient given small $m$ and initial SRS, especially given slowly changing networks, where the panel ACS does not yield much gain over the standard panel design. 
\end{itemize}
The improvements of iterated ACS is useful given relatively small $m$, if positively size-biased sampling is difficult to achieve, e.g. due to a lack of understanding of the relevant risk factors, or a lack of frame data that can be used to effectively vary the initial sample inclusion probability even though the relevant factors are known. Together, the panel and iterated ACS designs seem to complement each other in different settings, offering helpful choices across a wider range of situations than each on its own.

\appendix 
\section{Notes on sampling variances} 
 
Under ACS, let $\pi_{(i)} = \mbox{Pr}(i\in s)$ for individual $i$, let $\pi_{(\kappa)}$ be the sample inclusion probability of its network $\kappa$. Denote by $\beta_{\kappa}$ all the cases in the network $\kappa$. We have 
\[
\pi_{(i)} = \pi_{(\kappa)} = 1 - \bar{\pi}_{\beta_\kappa} 
\]
where $\bar{\pi}_{\beta_{\kappa}}$ is the exclusion probability of $\kappa$ under ACS, which is the probability that none of $\beta_{\kappa}$ is included in the initial sample $s_0$ under the sampling design of $s_0$. Let $\pi_{(ij)}$ be the joint sample inclusion probability of individuals $(i,j)$ under ACS, which is equal to that of the networks $\kappa$ and $\ell$ each of them belongs to, respectively. We have
\[
\pi_{(\kappa\ell)} = 1 - \bar{\pi}_{\beta_\kappa}  - \bar{\pi}_{\beta_\ell} + \bar{\pi}_{\beta_{\kappa}\cup\beta_{\ell}} 
\]
The variance of the HT-estimator of $\theta$ under ACS, derived from \eqref{ACS-HT}, is given by
\begin{align*}
V(\hat{\theta}) = \frac{1}{N^2} \sum_{\substack{i\in U\\ y_i=1}} \sum_{\substack{j\in U\\ y_j=1}} \big( \frac{\pi_{(ij)}}{\pi_{(i)} \pi_{(j)}} -1 \big) 
= \frac{1}{N^2} \sum_{\kappa =1}^M \sum_{\ell =1}^M  \big( \frac{\pi_{(\kappa\ell)}}{\pi_{(\kappa)} \pi_{(\ell)}} -1 \big) n_{\kappa} n_{\ell}
\end{align*}
where $M$ is the number of case networks in the population, since only the cross-products $y_i y_j$ of \emph{cases} $(i,j)$ contribute to the summation of $(i,j)$ over $U\times U$. 

Under the panel ACS design, the variance of each HT-estimator in $\hat{\nabla}_{t,t+1}^{pACS}$ follows from that of \eqref{ACS-HT} above. Similarly for the covariance between them, where we have
\begin{align*}
Cov(\hat{\mu}_t, \hat{\mu}_{t+1}) & = \frac{1}{N_t N_{t+1}} \sum_{\substack{i\in U_t\\ y_{i,t}=1}} \sum_{\substack{j\in U_{t+1}\\ y_{j,t+1}=1}} \big( \frac{\pi_{(ij)}}{\pi_{(i)} \pi_{(j)}} -1 \big) y_{i,t} y_{j,t+1} \\
& = \frac{1}{N_t N_{t+1}} \sum_{\substack{\kappa\in \Omega_t\\ n_{\kappa,t} >0}} \sum_{\substack{\ell\in \Omega_{t+1}\\ n_{\ell,t+1} >0}} \big( \frac{\pi_{(\kappa\ell)}}{\pi_{(\kappa)} \pi_{(\ell)}} -1 \big) n_{\kappa,t} n_{\ell,t+1}
\end{align*}
The first-order inclusion probabilities can be calculated as usual. For the second-order inclusion probabilities, notice that the two networks $\kappa$ and $\ell$ refer to two different time points here, such that one needs to take into account two population edge sets instead of only one, similarly as detailed below for iterated ACS design. 
 
Under the iterated ACS design, one can view the estimator \eqref{iACS} as a HT-estimator based on $s(t)$, with associated value $y_{i,t+1}/N_{t+1} - y_{i,t}/N_t$ for each $i\in s(t)$. The variance follows. The inclusion probability of $i\in s(t)$ has been explained before. Let $\kappa$ be the network of individual $i$, where $\beta_{\kappa} = \{ i\}$ if $y_{i,t} = 0$. Let $\ell$ be that of $j$. The joint inclusion probability of $i\neq j \in s(t)$ is given by
\[
\pi_{(ij)} = \begin{cases} \pi_{ij} & \text{if } y_{i,t} = 0, ~ y_{j,t} = 0 \\ 
\pi_i + \pi_{(j)} + \bar{\pi}_{\{i\} \cup \beta_{\kappa}} -1 & \text{if } y_{i,t} =0,~ y_{j,t} = 1 \\ 
\pi_{(i)} + \pi_j + \bar{\pi}_{\beta_{\kappa} \cup \{j\}} -1 & \text{if } y_{i,t} =1,~ y_{j,t} = 0 \\ 
\pi_{(i)} + \pi_{(j)} + \bar{\pi}_{\beta_{\kappa} \cup \beta_{\ell}} -1 & \text{if } y_{i,t} = 1,~ y_{j,t} = 1 \end{cases} 
\]

\end{document}